# Numerical Study of a Three-Dimensional Mixed Ising Ferrimagnet in the Presence of an External Field

G. M. Buendía and N. Hurtado

*Department of Physics, Universidad Simón Bolívar, Apartado 89000, Caracas 1080, Venezuela*

We present a numerical study based on a Monte Carlo algorithm of the magnetic properties of a mixed Ising ferrimagnetic model on a cubic lattice where spins $\sigma = \pm 1/2$ and spins $S = 0, \pm 1$ are in alternating sites on the lattice. We carried out exact ground state calculations and employ a Monte Carlo simulation to obtain the finite-temperature phase diagram of the model. A compensation point appears when the next-nearest-neighbor interaction between the spins $\sigma = \pm 1/2$ exceeds a minimum value. We found a strong dependence of the compensation temperature with the interactions in the Hamiltonian, particularly the crystal and the external field. An applied field can change the range of values of the compensation temperature from zero up to a maximum value that depends on the field.

## 1. Introduction

Ferrimagnetic alloys have been the object of intense experimental studies because their role in high-density magneto-optical recording [1, 2]. In a ferrimagnet the different temperature dependences of the sublattice magnetizations raise the possibility of the existence of compensation temperatures: temperatures below the critical point where the total magnetization is zero [3]. It has been shown that the coercivity is a function of the temperature and that it has a peak at the compensation point, favoring the creation of small, stable, magnetic domains [4, 5]. In magneto-optical recording devices the coercivity is changed by local heating of the media with a focused beam. Direct overwrite capability has been demonstrated in amorphous ferrimagnetic films with compensation temperatures higher than room temperatures [2, 6]. Also, the strong and continuous effort to synthesize low-density, transparent magnets, with spontaneous moments at room temperature, has reopened the interest in studying the ferrimagnetic ordering that plays a fundamental role in these materials [7 to 9]. In this work we describe a three-dimensional, spin 1/2–spin 1, mixed Ising model with ferrimagnetic ordering that, under certain conditions, exhibits compensation points. Mixed Ising systems have been studied before with mean-field [10, 11] and nonperturbative techniques [12 to 14] providing interesting information about the interactions responsible for the existence of compensation points. In the present study we extend previous Monte Carlo simulations of a two dimensional model to a three-dimensional case and include the effect of external fields. We are interested particularly in the ground state, the finite-temperature phase diagrams, and the location and characterization of the compensation temperatures.

## 2. The Model and Its Ground States

We study a three-dimensional Ising system with spins $\sigma = \pm 1/2$ and $S = \pm 1, 0$ located in alternating sites of a cubic lattice. A Hamiltonian for this model that includes nearest neighbor interactions between the $S$ and the $\sigma$ spins, next-nearest neighbor interactions between the $\sigma$ spins, crystal field and external magnetic field can be written as

$$\mathcal{H} = -J_1 \sum_{\langle nn \rangle} \sigma_i S_j - J_2 \sum_{\langle nnn \rangle} \sigma_i \sigma_k - D \sum_j S_j^2 - H \left( \sum_i \sigma_i + \sum_j S_j \right), \qquad (1)$$

where $J_1$ and $J_2$ are the exchange-interaction parameters, $D$ is the crystal field and $H$ is the external field, all in energy units.

The models to be considered will be labeled by enumerating the parameters different from zero in the Hamiltonian. For example the $J_1$–$D$ one is the model with all the parameters in the above Hamiltonian zero except $J_1$ and $D$. Since we are interested in studying the possible existence of compensation points, all our numerical results are obtained for the ferrimagnetic coupling, $J_1 < 0$, between nearest neighbors. We include only the next-nearest neighbor interaction, $J_2$, between the $\sigma$ spins because previous work done in two dimensional models shows that it is responsible for the existence of compensation temperatures [12, 13].

Since for our Hamiltonian, the ground state is translationally invariant, to obtain the ground state diagrams we only have to calculate the energy of the configurations of a $2 \times 2 \times 2$ unit cell [15]. This cell has $2^4 \times 3^4$ configurations. Taking into account rotational symmetries we found that there are 75 different configurations. The ground state of the model depends on the particular choice of parameters in the Hamiltonian, as can be seen in Figs. 1 and 2. In Fig. 1 we show the ground state diagram for the $J_1$–$J_2$–$D$ model ($H = 0$ and $J_1 < 0$), there are four different regions. The ground state diagram for the $J_1-J_2-D-H$ model ($J_1 < 0$, $H > 0$ and $D > 0$) is divided in six regions as seen in Fig. 2. The configurations of the unit cell in each region are labeled as in Table 1. The equations of the boundaries between regions are obtained by pairwise equating the ground-state energies.

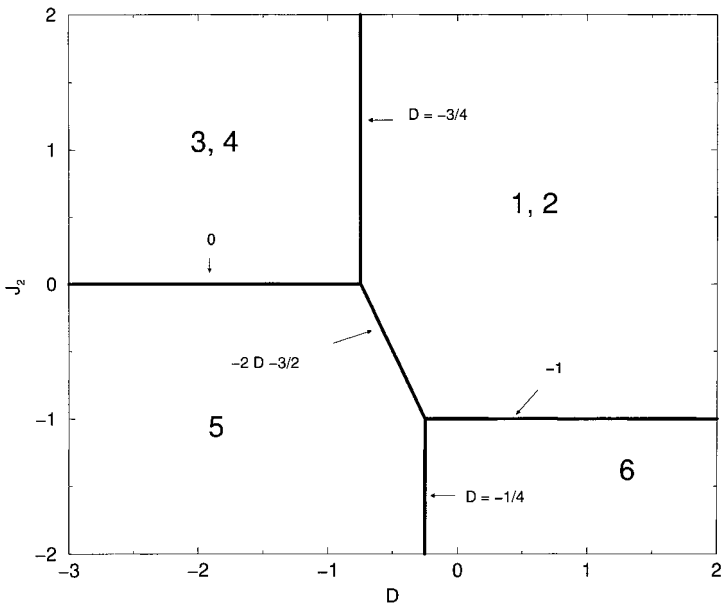

Fig. 1. Ground state diagram for the $J_1$–$J_2$–$D$ model ($J_1 = -1/2$). The configurations of the unit cell in each of the four regions are labeled as in Table 1

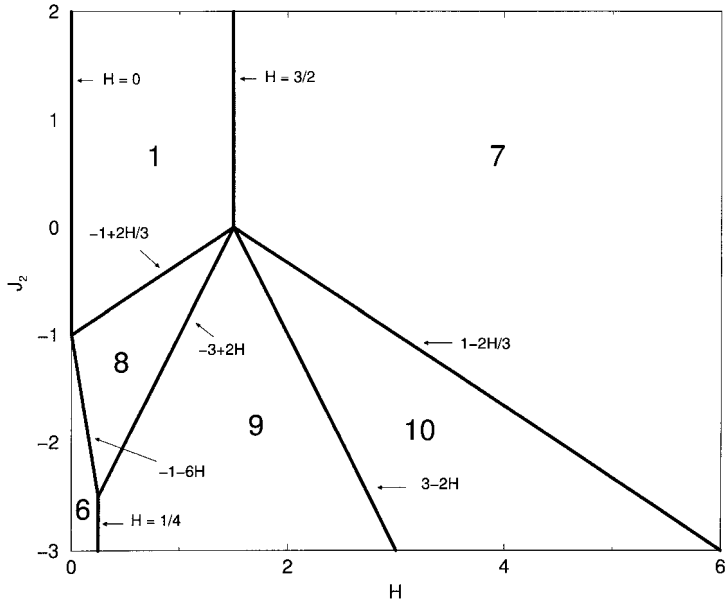

Fig. 2. Ground state diagram for the $J_1-J_2-D-H$ model ($J_1 = -1/2$, $D = 1$, $H > 0$). The configurations of the unit cell in each of the six regions are labeled as in Table 1

Table 1
Ground-state configurations, degeneracies and energies of the $2 \times 2 \times 2$ unitary spin cells indicated in Figs. 1 and 2. The symbol convention is: $S = \Uparrow, \Downarrow, 0$ and $\sigma = \uparrow, \downarrow$. The upper line gives the clockwise alignment of the spins at the top face of the cube, and the lower one the clockwise alignment at the bottom face

| | configuration | degeneracy | energy per site |
|---|---|---|---|
| 1 | $\Uparrow \downarrow \Uparrow \downarrow$ / $\downarrow \Uparrow \downarrow \Uparrow$ | 1 | $E = 6J_1 - \frac{3}{2}J_2 - 4D - 2H$ |
| 2 | $\Downarrow \uparrow \Downarrow \uparrow$ / $\uparrow \Downarrow \uparrow \Downarrow$ | 1 | $E = 6J_1 - \frac{3}{2}J_2 - 4D + 2H$ |
| 3 | $0 \uparrow 0 \uparrow$ / $\uparrow 0 \uparrow 0$ | 1 | $E = -\frac{3}{2}J_2 - 2H$ |
| 4 | $0 \downarrow 0 \downarrow$ / $\downarrow 0 \downarrow 0$ | 1 | $E = -\frac{3}{2}J_2 + 2H$ |
| 5 | $0 \uparrow 0 \uparrow$ / $\downarrow 0 \downarrow 0$ | 6 | $E = \frac{1}{2}J_2$ |
| 6 | $\Uparrow \downarrow \Uparrow \downarrow$ / $\uparrow \Downarrow \uparrow \Downarrow$ | 6 | $E = 2J_1 + \frac{1}{2}J_2 - 4D$ |
| 7 | $\Uparrow \uparrow \Uparrow \uparrow$ / $\uparrow \Uparrow \uparrow \Uparrow$ | 1 | $E = -6J_1 - \frac{3}{2}J_2 - 4D - 6H$ |
| 8 | $\Uparrow \uparrow \Uparrow \downarrow$ / $\downarrow \Uparrow \downarrow \Uparrow$ | 4 | $E = 3J_1 - 4D - 3H$ |
| 9 | $\Uparrow \uparrow \Uparrow \uparrow$ / $\downarrow \Uparrow \downarrow \Uparrow$ | 6 | $E = \frac{1}{2}J_2 - 4D - 4H$ |
| 10 | $\Uparrow \uparrow \Uparrow \uparrow$ / $\uparrow \Uparrow \downarrow \Uparrow$ | 4 | $E = -3J_1 - 4D - 5H$ |

## 3. The Monte Carlo Simulations

We used standard importance sampling methods [16], to simulate the model described by equation (1) on a cubic lattice of volume $L \times L \times L$ with periodic boundary conditions. Configurations were generated by sequentially traversing the lattice and making single-spin flip attempts at each site. The flips are accepted or rejected according to a heat-bath algorithm. One Monte Carlo Step (MCS) was defined as $L \times L \times L$ attempted spin moves, $10^4$ MCS were used to obtain each data point in lattices with $L = 40$, after discarding the first $10^3$ steps. We define $\beta = 1/k_B T$ and take Boltzmann's constant $k_B = 1$. Our program calculates the internal energy per site,

$$E = \frac{1}{L^3} \langle H \rangle, \tag{2}$$

the specific heat per site,

$$C = \frac{\beta^2}{L^3} [\langle H^2 \rangle - \langle H \rangle^2], \tag{3}$$

the sublattice magnetizations per site,

$$M_1 = \frac{2}{L^3} \left\langle \sum_i S_i \right\rangle, \qquad M_2 = \frac{2}{L^3} \left\langle \sum_j \sigma_j \right\rangle, \tag{4}$$

the total magnetization per site, $M = (M_1 + M_2)/2$, and the susceptibility,

$$\chi = \beta(\langle M^2 \rangle - \langle M \rangle^2). \tag{5}$$

In order to locate the compensation point, we define an order parameter per spin as

$$O = \frac{1}{L^3} \left\langle \left| \sum_{ij} (S_i + \sigma_j) \right| \right\rangle, \tag{6}$$

which is equivalent to the average of the absolute value of the total magnetization.

At the compensation temperature, $T_{\text{comp}}$, the sublattice magnetizations have equal magnitude and opposite signs

$$|M_1(T_{\text{comp}})| = |M_2(T_{\text{comp}})| \tag{7}$$

and

$$\text{sign}[M_1(T_{\text{comp}})] = -\text{sign}[M_2(T_{\text{comp}})], \tag{8}$$

such that the total magnetization is zero. The $T_{\text{comp}}$ is always lower than the critical temperature, $T_c$.

## 4. Results

### 4.1 Effect of next-nearest-neighbor interactions ($J_1$–$J_2$–$D$ model)

Recent results show that mixed Ising ferrimagnetic models in two dimensions present compensation points when the next-nearest-neighbor interaction between the $\sigma$ spins ($J_2$) is taken into account [12, 13]. We found that this is also the case for the three-dimensional model. The order parameter curves shown in Fig. 3 indicate the presence of compensation temperatures when $J_2$ exceeds some value that depends on the other parameters in the Hamiltonian. Notice that the compensation temperature remains basically unchanged once $J_2$ exceeds its minimum value; however, as expected, the critical temperature keeps increasing with increasing values of $J_2$. As can be seen in Fig. 4, the compensation temperature occurs because as the $J_2$ parameter increases, the ferromagnetic interaction between the $\sigma$ spins grows, such that the $\sigma$ sublattice remains ordered at higher temperatures. Since at the same time, the $S$ sublattice magnetization ($M_1$) is decreasing as the temperature is increasing, there is a temperature below the critical one at which both sublattice magnetizations (that have opposite signs) cancel each other and the compensation point occurs. Further increase of $J_2$ does not change the compensation temperature that has already been reached, but has the effect of keeping the system ordered at higher temperatures such that the critical point occurs at higher temperatures. In Fig. 5, we plotted the critical and compensation temperatures as a function of $J_2/|J_1|$ for a fixed value of $D/|J_1|$. The critical temperatures were obtained by locating the maxima of the specific heat. These plots show that the compensation temperature appears when the $J_2$ interaction takes some minimum value, that depends on $D/|J_1|$, after which $T_{\text{comp}}$ is almost independent of $J_2$.

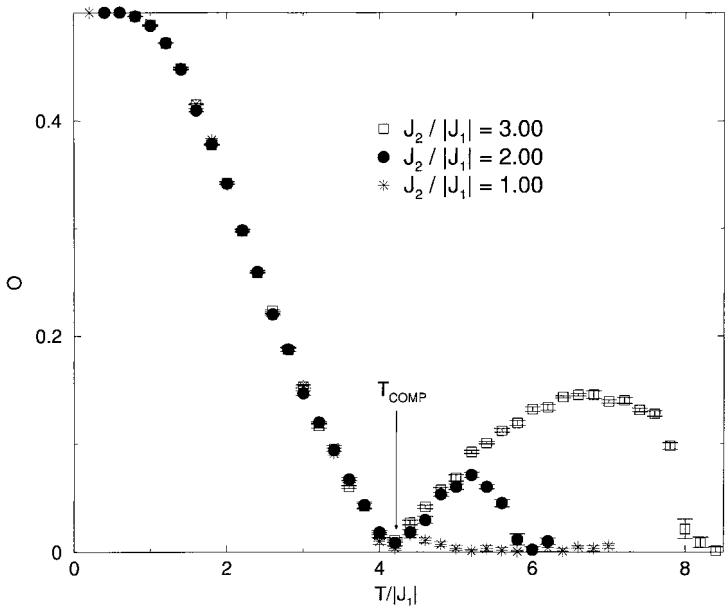

Fig. 3. Order parameter (as defined in equation (6)) vs. temperature for the $J_1$–$J_2$–$D$ model at $D/|J_1| = 2$ and several values of $J_2/|J_1|$. Notice that $T_{\text{comp}}$ is nearly independent of $J_2/|J_1|$ (for $J_2 < J_2^{\min}$), but $T_c$ increases with increasing values of $J_2/|J_1|$

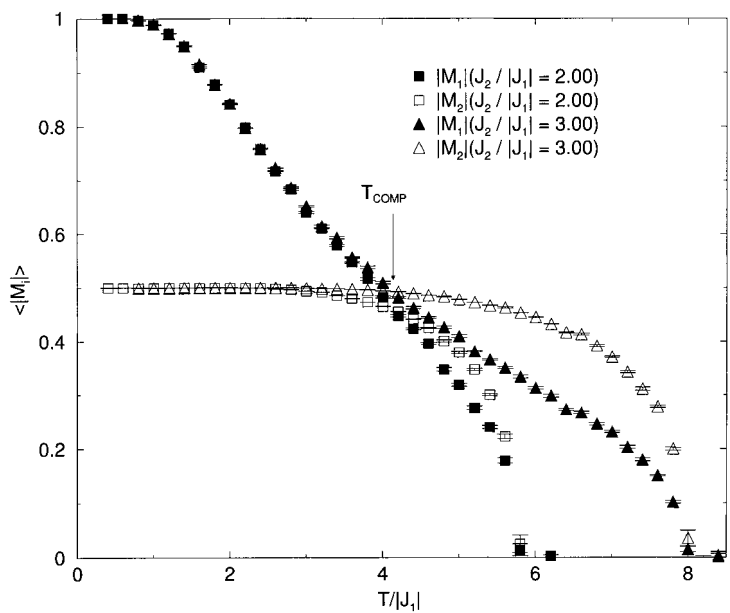

Fig. 4. Absolute value of the sublattice magnetizations vs. the temperature for the $J_1-J_2-D$ model with $D/|J_1|=2$. The $T_{\text{comp}}$ is reached at the crossing point $|M_1(T_{\text{comp}})|=|M_2(T_{\text{comp}})|$ where the order parameters is a minimum (as seen in Fig. 3)

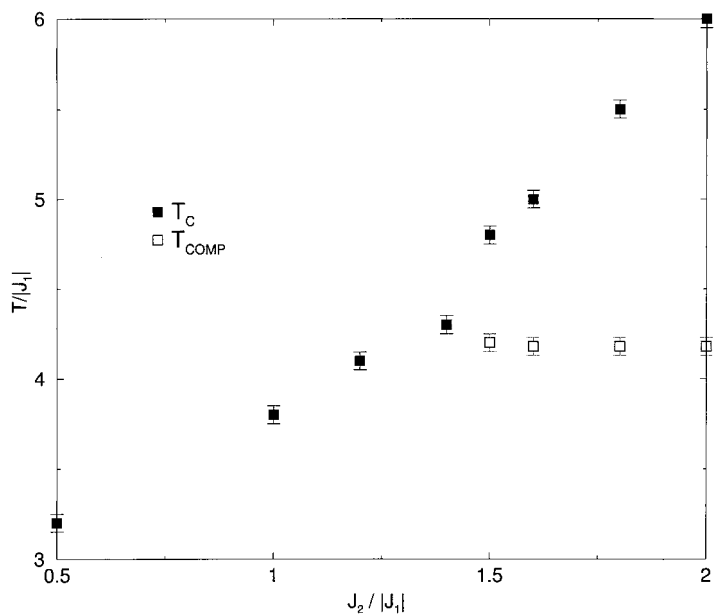

Fig. 5. Critical and compensation temperature vs. $J_2/|J_1|$ $(D/|J_1|=2, H/|J_1|=0)$

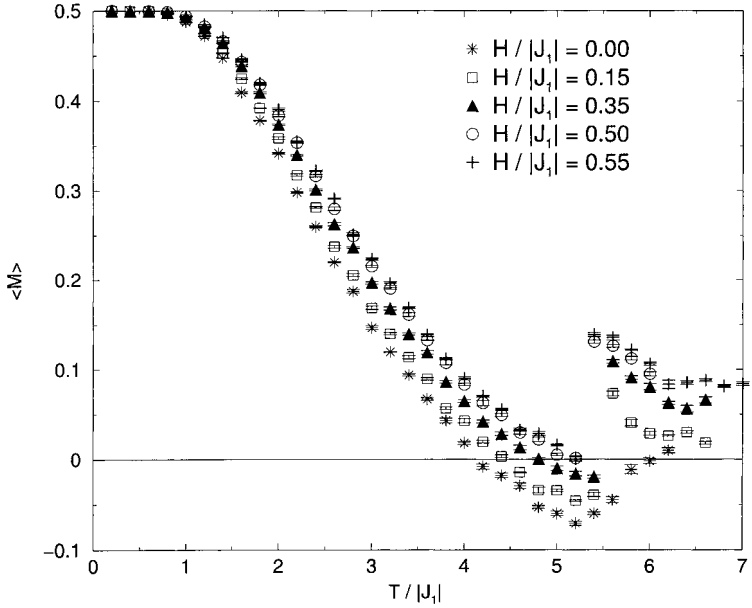

Fig. 6. Total magnetization vs. temperature for the $J_1-J_2-D-H$ model with $J_2/|J_1|=2$ and $D/|J_1|=2$, for several values of $H$. For this choice of parameters the compensation point disappears for $H/|J_1|>0.55$

### 4.2 Effect of external magnetic fields ($J_1-J_2-D-H$ model)

In this section we are going to study the behavior of the mixed Ising ferrimagnetic model in the presence of external magnetic fields. For this study we must be very careful selecting our initial connfiguration. At low temperatures the system has metastable states with very long lifetimes (longer that our Monte Carlo runs). In order to calculate the compensation temperature, we start our measurements once the system has been warmed from its ground state.

In Fig. 6 we plot the total magnetization as a function of the temperature for different values of $H/|J_1|$ and a fixed value of $J_2/|J_1|$ and $D/|J_1|$. Notice that the compensation temperature increases with the field, but the temperature at which the total magnetization becomes discontinuous, $T_D$ decreases with it, until they become equal for a certain value of $H$. For higher values of $H$ the system does not have a compensation temperature. This limit value of $H$ depends on the other parameters in the Hamiltonian.

In Fig. 7 we plot $T_{\text{comp}}$ versus $H/|J_1|$ for a couple of values of $D/|J_1|$, notice that $T_{\text{comp}}$ increases with the value of $D/|J_1|$. In general we see that $D$ and $H$ have the similar effect of increasing $T_{\text{comp}}$, but $T_D$ seems to be less sensible to changes in $D$. The effect of both, the crystal field and the external field, is to change the temperature from a range that varies from zero up to $T_D$.

### 5. Conclusions

We have applied a Monte Carlo algorithm to simulate a mixed Ising system on a cubic lattice, where spins, $\sigma=\pm 1/2$ are alternated with spins $S=\pm 1,0$. The Hamilto-nian includes nearest-, and next-nearest-neighbor interactions, crystal field and external field. By choosing an antiferromagnetic nearest-neighbor interaction we have a simple but interesting model of ferrimagnetic ordering. We have calculated the ground-state energies and the finite-temperature phase diagram showing the compensation and critical temperatures. Our results show that, as happens in the two dimensional case [12, 13], the compensation temperatures appear when the next-nearest neighbor interaction between the $\sigma$ spins, $J_2$, is included. The minimum value of $J_2$ for a compensation point to appear depends on the value of the other parameters in the Hamiltonian. In absence of external field ($J_1-J_2-D$ model) the compensation temperature remains approximately constant for any value of $J_2$ above the minimum and for a fixed value of $D$. The effect of the external field is to change the compensation temperature in a range that varies from zero up to the value at which the total magnetization becomes discontinuous, $T_D$. The compensation temperature increases in a way that seems linear with the field until it disappears for strong enough fields. Since the compensation point is not a critical point we expect that finite-size effects are not relevant for this study, except in the region where $T_{comp}$ approaches $T_D$ that is a size dependent quantity. So, finite size effect can change the range of values of the compensation temperature.

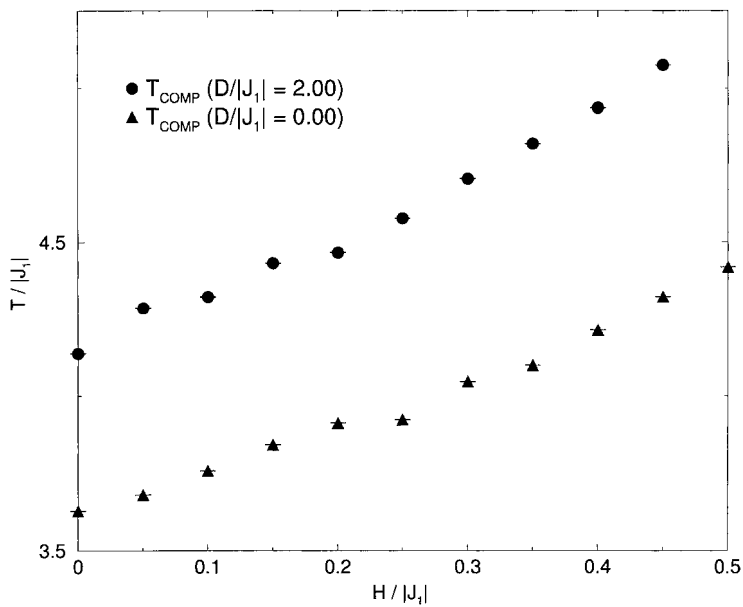

Fig. 7. Compensation temperatures vs. $H$ for the $J_1-J_2-D-H$ model at $J_2/|J_1|=2$ and $D/|J_1|=0$ and 2

Our study suggests that there is a strong dependence between the compensation temperature and the interactions in the Hamiltonian. Not only the value of the compensation temperature depends on the parameters but even its existence. An external magnetic field can be used to change the range of values of the compensation temperature. Experimental evidence of the dependence on compensation points with the interactions has already been found [17, 18].